\documentclass[%
 reprint,
 amsmath,amssymb,
 aps,
]{revtex4-2}
\usepackage{comment}
\usepackage{graphicx}
\usepackage{dcolumn}
\usepackage{bm}

\begin{document}

\preprint{APS/123-QED}

\title{Maximal energy extraction through information gathering}

\author{Michael Grayson}
 \altaffiliation{Department of Electrical, Computer and Energy Engineering, University of Colorado, Boulder, CO 80309, USA}

\author{Charlie Rackson}
\affiliation{Department of Electrical, Computer and Energy Engineering, University of Colorado, Boulder, CO 80309, USA}



\date{\today}

\begin{abstract}
We calculate the maximal non-equilibrium work that can be extracted from any mass using information about its micro-state. This is done through the use of black hole thermodynamics in the context of the thermodynamics of information. The non-equilibrium work that can be obtained by knowing the exact micro-state is found to be $\frac{1}{2} Mc^{2}$. This is calculated using the entropy of an eternal, uncharged, non rotating black hole. This is derivation helps elucidate the relativity of entropy and its relationship to general relativity and time. 


\end{abstract}

\maketitle


\section{\label{sec:level1} Introduction}

In the classical thermodynamics, entropy is treated as an absolute thermodynamic property: one that is not relative to the observer \cite{schroeder_introduction_1999}. Through the unification of classical and quantum mechanics, it has been determined that entropy is a quantity that is relative to the observer and how much information about the system in question they have \cite{parrondo_thermodynamics_2015}. This information can be thought of as entanglement: the sharing of correlations between the observer and the observed. In a classical framework, this shared information or correlation is known as mutual information \cite{parrondo_thermodynamics_2015}. The amount of mutual information an observer shares with a system, in addition to the free energy, places an upper bound on the amount of work an observer can extract from that system \cite{parrondo_thermodynamics_2015}. This work can be extracted from the system by the observer as work through feedback control \cite{sagawa_second_2008}. This feedback control process has a time reversal that corresponds to the energy dissipation required for information storage and deletion \cite{kawai_dissipation_2007, rubino_inferring_2022}. These recent theoretical principles have revolutionized our understanding of the arrow of time and the meaning of information within physical theories. However, these formulations are still very new and their implications have not been extensively studied. In particular, the experienced passage of time is not a continuous parameter, but an ensemble average susceptible to fluctuations \cite{rubino_quantum_2021}. This ensemble average rate of time will increase with entropy production in the observers frame and decrease with information \cite{parrondo_thermodynamics_2015}. This newly discovered relativity of entropy and time requires reconsideration of the usage of time in other areas of physics such as general relativity.

Modern gravitational theory was revolutionized by the union of thermodynamics and gravitational theory \cite{wald_thermodynamics_2001, bousso_holographic_2002, bekenstein_black_2004}.  These revolutions began with the notion that black holes should behave as a thermodynamic system with an entropy proportional to the event horizon surface area \cite{bardeen_four_1973}. Additionally, it was shown that a black hole should radiate black-body radiation due to the interaction of vacuum fluctuations at the event horizon and gravitational red-shifting \cite{hawking_particle_1975}. It was then realized that black holes not only behave as a thermodynamic system, but they also act as upper bounds to what a thermodynamic system can be. These bounds include the maximum entropy per volume of space time and the rate at which information can be scrambled \cite{bekenstein_black_2004, sekino_fast_2008}. In its simplest formulation, a black hole has the maximal entropy that any volume of spacetime can have \cite{bekenstein_universal_1981}. This entropy is proportional not to the volume of the black hole, as expected classically, but to its surface area. This drastically differs from the thermodynamic notion that the entropy of an object is simply proportional to the number of atoms in an object (standard molar entropy). The unification of black hole thermodynamics and classical thermodynamics perhaps involves the relativity of entropy to the observer \cite{parrondo_thermodynamics_2015}. Black hole thermodynamics has also lead to insights into how a quantum theory corresponds to the boundary of a gravitational theory \cite{hubeny_adscft_2015,bousso_holographic_2002, susskind_world_1995}. This principle is known as Ads-CFT correspondence, and has become the foundation of insights into the quantum nature of gravity \cite{hubeny_adscft_2015}. Ads-CFT correspondence equates the entropy of a 3 dimensional volume of space with the entropy of a 2 dimensional quantum theory at its surface. This yields the unexpected scaling of entropy with surface area and not volume in gravitational theories. In particular, entropy has been found to be related to the structure of spacetime and the behavior of black holes \cite{hubeny_adscft_2015, susskind_teleportation_2018}. Through the Ryu-Takanagi conjecture - an extension of black hole entropy - it is expected that the curvature of space is related to the entanglement entropy between regions of space \cite{ryu_aspects_2006}.

In this paper we answer the simple question of: given a mass $M$ how much energy can be extracted if we know the exact micro-state of the mass? More specifically, if we know all the information that it is possible to know about a mass $M$, how much energy can we extract from it using information alone? In a perfect formulation this would include every degree of freedom possible for a system. In this derivation, we consider the degrees of freedom accounted for by black hole thermodynamics. To put it more succinctly, this paper answers the question: \emph{What is the maximum energy can be extracted from a mass using information alone?}

\section{\label{sec:level2} Thought Experiment}

\begin{figure}[htbp]
\includegraphics[height=4.5cm]{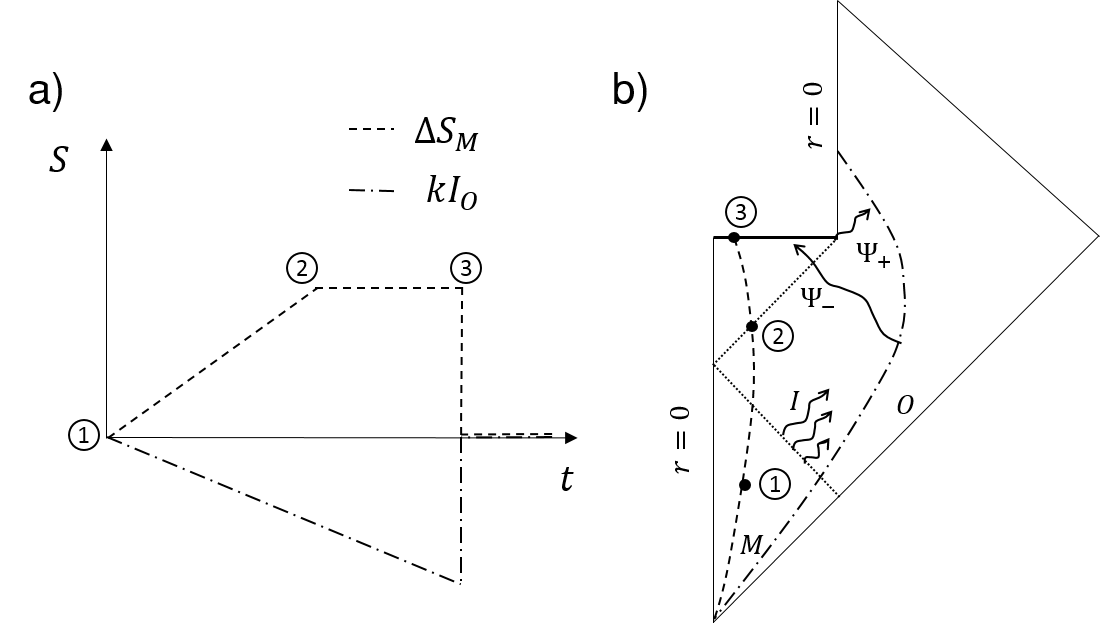}
\caption{\label{fig:epsart} a) a depiction of the entropy as a function of time for the matter and the observer. The entropy of the mass is shown as a dashed line, while the information of the observer is a dashed dotted line. the matter increases in entropy as it collapses into a black hole $1\rightarrow 2$, then remains constant $2 \rightarrow 3$. Finally the observer perturbs the black hole to reduce its entropy while using up their information. b) a Penrose diagram of the process with corresponding time slices label to correpond to the events in a). The mass is shown as a  dashed line while the observer is shown as a dotted line. The event horizon is shown as a dotted line and flows of information or perturbations are squiggly lines.}
\end{figure}

We will answer this question by using a thought experiment in which we collapse some mass $M$ into black hole to force it into its maximal entropy state. We can then calculate the entropy and temperature of this black hole using the well known thermodynamic equations for a black hole. Finally we will calculate the information required to specify the micro-state of this black hole and the work that can be obtained through some kind of feedback control. Figure \ref{fig:epsart} (a) depicts the entropy of the mass as well as the information of the observer. The observer collects information from the mass, likely photons, as the mass collapses and increases in entropy. The mass then passes fully into the event horizon while the observer continues to collect information. Finally the observer applies some kind of perturbation to the black hole based on the information they have which causes the black hole to evaporated a large amount of energy. This perturbation acts much like hawking radiation except that it is specifically prepared by the observer. Figure \ref{fig:epsart} (b) depicts the Penrose diagram for this process with the number circles corresponding to those in (a). initially there is a mass at rest and an observer at rest some radius $r$ from the mass this mass collapses into a black hole. The observer collects information encoded into the null boundary that corresponds to the black hole horizon extended into the past \cite{moustos_gravity_2017}. After the observer collects sufficient information they apply the perturbation $\Psi_{-}$ and collect emitted radiation $\Psi_{+}$. In general this perturbation $\Psi_{-}$ can be adiabatic and require no work to perform \cite{parrondo_thermodynamics_2015, sagawa_second_2008}. Perhaps it is possible through the interaction of quantum fields with hawking radiation at the horizon of the black hole. To perform this operation the observer would need to use past information to compute the current micro-state of the black hole at to produce $\Psi_{-}$. This process would require a computation system at least as complex as the black hole itself by the law of requisite variety \cite{ashby_requisite_1991}. This computation would necessarily use energy due to Landauer's Principle \cite{kawai_dissipation_2007}.

To derive the energy that can be extracted through a mass $M$ through information we will need assume that the maximal entropy state of any system is a black hole. This is a valid assumption as a black hole produces the maximum entropy density that can be held within a certain volume of space-time \cite{bekenstein_universal_1981}. This maximum entropy is determined by the surface area of the event horizon of a black hole and is known as the Bekenstein bound\cite{bekenstein_universal_1981}. In modern black hole thermodynamics a black hole can be though of as a perfect scrambler of information \cite{brown_complexity_2016, bouland_computational_2019}. As objects fall into a black hole their information is rapidly spread through the black hole. In fact it is spread at the maximal rate making a black hole a perfect scrambler \cite{brown_complexity_2016, bouland_computational_2019}. We can begin to imagine that our model for a black hole merely consists of the absolute limit of a complex system. Systems that rapidly spread correlations such as chaotic systems are a black hole with slower scrambling times. That is why a black hole acts as an upper bound to entropy density.

We will also need to assume that it is possible to create a black hole from any mass $M$. While we know large black holes can form it is not necessarily true that small mass black holes can form. We could try to get around this by lowering a mass into a black hole that has already formed. The black hole mass and entropy will be $M+\Delta m$ which will give a entropy $S+\Delta S$ where $\Delta S \propto M\Delta m+\Delta m^{2}$. Therefore there are contributions to the entropy due to the mixing of the mass $M$ and $\Delta m$ we would like to ignore. We will prove that spontaneous collapse will occur in a finite time for any mass $M$ by employing the fluctuation theorem. The observer will then need only to wait for a sufficiently long time to observer a collapse.

\subsection{Spontaneous collapse of matter}

First we will show using the fluctuation theorem that an arbitrary mass $M$ will eventually become a black hole. To due this we will use Crook's fluctuation theorem. The fluctuation theorem is a very general theory but does have its own assumptions that may be interesting to note. It requires that the underlying physics be time reversible, ergodic and causal \cite{crooks_entropy_1999, parrondo_thermodynamics_2015, evans_fluctuation_2002}. Time reversibility will be generally true unless weak nuclear forces dominate at some point during the evolution. We will assume this is not the case. Ergodicity may or may not be true. The assumption of ergodicty requires that there is a nonzero probability that the mass $M$ can become a black hole spontaneously. We think the validity of this assumption is heavily dependent on the size of the mass. In particular, we know very large masses can become black holes spontaneously, but it is unclear whether there is a mechanism on large time scales that can do this for small masses \cite{lodato_supermassive_2006}. We assume that this is the case to simplify the derivation. Finally we will require that causality is true. It is this assumption that may be inherently related to the perception of useful work and dissipated work as transitions in time.
Crook's fluctuation theorem relates the probability of a transition with its time reversal and the entropy produced in that reversal \cite{crooks_entropy_1999, evans_fluctuation_2002}. We assume that the entropy of the initial state is dominated by the black hole entropy. We label that state of ordinary mass as $A$ and that of the black hole as $B$. The fluctuation theorem is then written as
\begin{equation} \label{eq:fluc1}
\frac{P(A\rightarrow B)}{P(B \rightarrow A)} = e^{\Delta S}
\end{equation}
We assume that the transition from black hole to mass corresponds to Hawking radiation \cite{hawking_black_1974}. The entropy of the black hole is larger than that of the original mass. so we can conclude that $\exp(\Delta S)>1$ and that 
\begin{equation} \label{eq:fluc2}
P(A\rightarrow B)>P(B \rightarrow A) 
\end{equation}
Assuming the times scales for both transitions are equal, the lifetime of the black hole state will be inversely proportional to the probability of its transition to mass $\tau_{evap} \propto 1/P(B\rightarrow A)$ and $\tau_{collapse} \propto 1/P(A\rightarrow B)$. We can additionally use the evaporation time as an upper bound.
\begin{equation} \label{eq:fluc3}
\tau_{collapse}<\tau_{evap}
\end{equation}
If we know the time scales over which the hawking radiation occurs we could calculate the actual rate of collapse using the black hole entropy and the hawking evaporation probability. If we assume that the transitions that are occurring is a small addition or subtraction of mass $\delta m$ over some small time $\delta t$ we can begin to calculate the exact rates of collapse from those of evaporation. We can convert the probabilities to transition rates by $P(M+\delta m)=\Gamma_{+}\delta t$ \cite{andrieux_fluctuation_2006, esposito_fluctuation_2006}
\begin{equation} \label{eq:fluc4}
\frac{P(M+\Delta M)}{P(M-\Delta M)} = e^{\frac{\Delta S}{\Delta t} \Delta t}
\end{equation}
\begin{equation} \label{eq:fluc5}
\frac{\Gamma_{+}}{\Gamma_{-}} = e^{\frac{\Delta S}{\Delta t} \Delta t}
\end{equation}
If we assume that the quanta emitted by the block hole have an average energy of $E=\hbar\omega$ and that each quanta is radiated over a lifetime of $\tau_{evap}=1/\Gamma_{-}$ then we have that the rate of change of the mass of the black hole is
\begin{equation} \label{eq:md1}
\dot{M}= \frac{\hbar \omega \Gamma_{-}}{c^{2}}
\end{equation}
One can calculate the tunneling rate of the hawking radiation as \cite{parikh_hawking_2000}
\begin{equation} \label{eq:fluc6}
\Gamma_{-} = e^{-8\pi\omega(M-\omega/2)}=e^{+\Delta S}
\end{equation}
therefore
\begin{equation} \label{eq:fluc7}
\Gamma_{+} =e^{+2\Delta S}=e^{-4\pi\omega(M-\omega/2)}
\end{equation}
This corresponds to a Boltzmann factor for a particle with energy $\omega$ at the inverse Hawking temperature $4\pi M$ \cite{parikh_hawking_2000}.
Which, as we expect, is the same as the evaporation transition rate, but squared. In the ensemble time average we would expect that the evaporation rate is equal to the condensation rate (non equilibrium partition identity). Therefore we would expect black holes to spontaneously form and this formation rate is exponentially proportional to its mass. We would therefore expect only very small small black holes to form spontaneously just as smaller black holes should radiate more strongly \cite{parikh_hawking_2000}. It also implies that a very large black hole is unlikely to accept a new particle unless the energy is very small which is counter to what we would expect from general relativity. In any case we have that form some arbitrary mass $M$ there is a probability of collapsing that is greater than the probability of evaporation. Therefore we'd expect over a long period of time that a arbitrary mass $M$ will collapse into a black hole and the the black hole is the likely for any mass $M$.



\subsection{Non-equilibrium work of a black hole}

The entropy of a system represents the amount of information required to specify the micro-state of the system. Conversely it represents the total amount of information missing from our current knowledge about the system. In general it increases with time as things become more disordered but it does have upper bounds. The thermodynamic entropy is defined as \cite{schroeder_introduction_1999} \[\int^{T}_{0} \frac{C_{p}}{T} dT\] which relates to how the degrees of freedom of a system are filled with thermal energy as one increases the temperature. Analytical expressions for the temperature dependence of $C_{p}$ exist, but usually the entropy of materials is an experimentally determined quantity \cite{schroeder_introduction_1999}. The free energy of a thermodynamic system is given by  \[F=H-TS \] where $F$ is the free or "usable energy", $H$ is the enthalpy or "total energy". \cite{parrondo_thermodynamics_2015} $TS$ corresponds to the energy lost when extracting work from $H$ due to the disorder of the system. If we gain information about the system we can reduce the amount of energy loss to disorder. We can relate the amount of free energy gained from the acquisition of information by \[F=H-TS+kTI\] where $I$ is the information we have about the system. This process of obtaining free energy from information is not reversible so information is used up to some extent to gain energy \cite{parrondo_thermodynamics_2015}. The maximum amount of information one can gain about the system corresponds to the total disorder of entropy of the system. This means that $kTI$ is bounded above by $TS$. For water at $25^o C$ $H\approx 1.9 \frac{MJ}{kmol}$ and $TS\approx 2.1 \frac{MJ}{kmol}$ meaning we could double the free energy by obtaining enough information to specify the microstate which would require $5e23$ $bits$ or 100 billion terabytes per kmol. This entropy is thermodynamic and only really accounts for kinetic degrees of freedom, many other degrees of freedom exists: electromagnetic, quantum mechanical, atomic, etc... It is this relativity of entropy and free energy that is suspiciously similar to the thermodynamics of quantum gravity. If entropy is relative to the observer and determines the curvature of space time two observers would experience different curvatures of space time. Alternatively the relatively of entropy may be isomorphic to the relativity one experiences in general relativity. We will analyze the situation where the entropy includes all possible degrees of freedom or at least the ones accounted for by black hole thermodynamics

The question is again: how much information can be obtained about some mass $M$ and how much energy can be converted to work through this information gathering proces?. We have shown that if an arbitrary mass $M$ is perfectly ergodic it will at some point be in a black hole. The black hole state has the maximum entropy per mass and will therefore give us the maximum amount of information that can be obtained and utilized about any mass $M$. The information energy of a black hole is given by \cite{parrondo_thermodynamics_2015} \[ TI=T_{b}S_{b}\] if we assume that the black hole is non rotating, and has no charge we obtain \cite{bekenstein_black_2004}
\[S_{b}={c^3 A\over 4 G \hbar}\]
\[A=16\pi(GM/c^2)^2\]
\[T_{b} ={ \hbar c\over 2k\pi}  {\sqrt{(GM/c^2)^2}\over r_h^2} \]
\[r= GM/c^2+\sqrt{(GM/c^2)^2}.\]
simplifying we obtain
\[T_{b}S_{b} = \frac{1}{2} Mc^{2}\]
We can additionally calculate the amount of information that would be required to do this which is 
\[I = \frac{4\pi GM^2}{\hbar c}\]
This seems somewhat obvious after we have derived it. the total amount of work we can obtain from a mass $M$ through information gathering is related the rest mass energy $E=Mc^{2}$. Why this additional term of $1/2$ though? If we also consider the white hole solution to contribute mass energy we obtain the total $kTI=\frac{1}{2} M_{bh}c^{2}+\frac{1}{2} M_{wh}c^{2}$. Why is is that only half the mass energy comes from the black hole? Experimentally it seems likely that white holes do not even exist in any straightforward way and are merely the time reversed solution to the black hole. Perhaps due to the relativity of entropy and the arrow of time both solutions to the black hole exist in the same space for different observers? An additional issue is that the quantity of information to do this is immense. One would have to be a black hole themselves to store the information. By the law of requisite variety if an observer wished to control this system they would need a higher entropy than $S_{b}(M)$ \cite{ashby_requisite_1991}. In Figure \ref{fig:epsart} the observer was required to predict the microstate of the black hole using some amount of computation. Modeling the black hole as a quantum computer the operations required for collapse are proportional to $log(S)$ while the energy extracted is on order $S$ \cite{sekino_fast_2008}. Assuming a black hole represents a lower bound on the energy per computation rate, the amount of energy used to predict the microstate by the observer is at minimum the energy radiated through hawking radiation by the black hole which will be negligible. Based on classical thermodynamics the perturbation can be adiabatic \cite{parrondo_thermodynamics_2015}. Therefore, the loss of energy relative to the rest mass energy must be due to the internal properties of the black hole. 

We can additionally perform the derivation in the presence of charge and angular momentum, both of which serve to reduce the total available free energy from information. However they do this by actually reducing the entropy of the black hole itself thereby increasing the total free energy. In equilibrium the density of trajectories for the Ford reverse process are equal and so in the production of work from equilibrium using only information we must
\[T_{b}S_{b} = \frac{c^4}{2 G} \sqrt{\frac{G^2 M^4-G M^2 Q^2-c^2 J^2}{c^4 M^2}}\]
The energy that can be extracted through information becomes zero when 
\[M=\sqrt{\frac{Q^2}{2G}-\frac{\sqrt{4 c^2 J^2+Q^4}}{2G}}\]
In this case the free energy of the black hole is now maximum and the work need not be extracted through some kind of feedback control, but can be extracted directly. For example, it is well known energy can be extracted by a rotating black hole through acceleration by a mass in its ergo-sphere. This may suggest that extremal black holes, ones which have large charge or momentum compared to there mass do not decay by emitting radiation but by producing work \cite{rosa_extremal_2010}.

\section{Discussion}
Work is produced from information by effectively reducing the relative entropy or Kullback-Leibler divergence between a forward process and the reverse process of some transition between equilibrium states with a feedback protocol during the transition \cite{sagawa_second_2008}. If a system has equal probability distributions in phase space for the forward and reverse processes there will be no work produced. If the probability distributions are not identical there will be dissipation in one time direction and work production in the reverse \cite{kawai_dissipation_2007}. Using information and feedback control one can modify the relative entropy between the forward and reverse processes producing work \cite{sagawa_second_2008, rubino_quantum_2021}. The observation of work produced in a dissipative system can be thought of as a projection onto its time reversal counterpart \cite{rubino_quantum_2021}. We can imagine that the production of $1/2Mc^{2}$ amount of work would imply that half of the systems total energy can be projected onto work while half of the systems energy can be projected onto dissipation. This is corroborated by the fact that the total mass energy $Mc^{2}$ has contributions equally from the white hole and black hole solutions. Therefore in an arbitrary mass $M$ that has a constant entropy there must a mixture of processes that have a forwards time direction (entropy production) and a backwards one (entropy reduction). This mixture must be equally balanced, half and half, to prevent entropy production or reduction. It is possible through some adiabatic control mechanism to perturb these systems to decouple them from one another, such that we can object half of the total mass energy of the object. 

A simple way to think of the result of $\frac{1}{2}Mc^{2}$ is in terms of matter and antimatter. If we wanted to extract the total energy from an object we could perform the following process. first we gather all possible information about the object. Then using this information we construct an object made antimatter such that the two will perfectly annihilate. For every perturbation of the degrees of freedom for the normal object we would need an equal and opposite one. This process would require the input of $mc^{2}$ antimatter and would release $2mc^{2}$ in energy. if we consider the total mass to be the combination of matter and antimatter we can assume $M=2m$. Then we will obtain a net release of $\frac{1}{2} Mc^{2}$. It is interesting that in both cases (black hole and matter-antimatter) we require half of the the system to be in a time reversed solution. Additionally, the time reversed object would likely not just be an antimatter version of the original if the object is very complex.

We could consider the total mass of the universe. If we had all the information of the universe we could only be able to extract $\frac{1}{2}M_{u}c^{2}$ from it. This would imply that in the universe half of the mass will correspond to a black hole solution and half to a white whole solution. This suggest that in the universe there is a type of conservation. That for any forwards time process there must be a backwards time process which can annihilate the original. we suspect that the relation between such systems should not be completely obvious. That that any forwards time process theoretically has a reverse, but not that this must exist in the universe in equal quantities. For example an electron traveling at some velocity has a time reversed solution which is a positron traveling at the same velocity in the opposite direction. However, it is not required that this particle should exist. It may exist, perhaps it is that we can negate our assumed direction of time to induce it to exist. There are however positrons in the universe. Many of these positrons are bound to neutrons in the form of protons. Something which does not occur for electrons. The universe is predominantly neutral in charge so there is likely an equal amount of positrons and electrons, an equal amount of time reversed and forwards solutions, but the reversed time solutions have been bound up in another process and no longer directly annihilate. With information we could unravel all these strings of complexity. Reducing every system into its forward and reversed components, and annihilate them all. They must be arranged to do so though, resulting in only half the energy being gained.

Perhaps we can generalize that any free energy that exists in the universe exists through this mechanism. That is through the coupling of systems who are time duals to each other. In the natural world we are constantly taking advantage of sources of free energy: solar radiation, stored chemical energy, thermal energy from the earths core, etc... If any source of free energy is due to the coupling of a system with its time reversed dual then we would expect that time reversed systems are just as common place in the natural world as their forward time partners. The process by which energy is stored then is due to the breaking up of a reversible process into a forwards and backwards time solution or in other words a symmetry breaking of time. The process of free energy production is the recombination of two time dual processes. 

\section{Conclusion}
 We calculated the rate at which a arbitrary mass $M$ will transition into a black hole. We then showed that the maximum amount of free energy that can be extracted from a mass $M$ using only information about that system is $\frac{1}{2}Mc^{2}$. The quantity is only half of the rest mass due to the contribution of the white hole solution to the back hole entropy. We further explained the factor of $1/2$ by considering the contributions from systems which spontaneously reduce and produce entropy. Since the direction of time is determined by the sign of entropy production \cite{rubino_inferring_2022} these are essentially forward and reverse time processes which coupled to produce net zero entropy production. We conjecture that any occurrence of free energy must be accompanied by the interaction of time reversal duals.



\bibliography{references}

\begin{thebibliography}{29}%
\makeatletter
\providecommand \@ifxundefined [1]{%
 \@ifx{#1\undefined}
}%
\providecommand \@ifnum [1]{%
 \ifnum #1\expandafter \@firstoftwo
 \else \expandafter \@secondoftwo
 \fi
}%
\providecommand \@ifx [1]{%
 \ifx #1\expandafter \@firstoftwo
 \else \expandafter \@secondoftwo
 \fi
}%
\providecommand \natexlab [1]{#1}%
\providecommand \enquote  [1]{``#1''}%
\providecommand \bibnamefont  [1]{#1}%
\providecommand \bibfnamefont [1]{#1}%
\providecommand \citenamefont [1]{#1}%
\providecommand \href@noop [0]{\@secondoftwo}%
\providecommand \href [0]{\begingroup \@sanitize@url \@href}%
\providecommand \@href[1]{\@@startlink{#1}\@@href}%
\providecommand \@@href[1]{\endgroup#1\@@endlink}%
\providecommand \@sanitize@url [0]{\catcode `\\12\catcode `\$12\catcode
  `\&12\catcode `\#12\catcode `\^12\catcode `\_12\catcode `\%12\relax}%
\providecommand \@@startlink[1]{}%
\providecommand \@@endlink[0]{}%
\providecommand \url  [0]{\begingroup\@sanitize@url \@url }%
\providecommand \@url [1]{\endgroup\@href {#1}{\urlprefix }}%
\providecommand \urlprefix  [0]{URL }%
\providecommand \Eprint [0]{\href }%
\providecommand \doibase [0]{https://doi.org/}%
\providecommand \selectlanguage [0]{\@gobble}%
\providecommand \bibinfo  [0]{\@secondoftwo}%
\providecommand \bibfield  [0]{\@secondoftwo}%
\providecommand \translation [1]{[#1]}%
\providecommand \BibitemOpen [0]{}%
\providecommand \bibitemStop [0]{}%
\providecommand \bibitemNoStop [0]{.\EOS\space}%
\providecommand \EOS [0]{\spacefactor3000\relax}%
\providecommand \BibitemShut  [1]{\csname bibitem#1\endcsname}%
\let\auto@bib@innerbib\@empty
\bibitem [{\citenamefont {Schroeder}(1999)}]{schroeder_introduction_1999}%
  \BibitemOpen
  \bibfield  {author} {\bibinfo {author} {\bibfnamefont {D.~V.}\ \bibnamefont
  {Schroeder}},\ }\href@noop {} {\emph {\bibinfo {title} {An {{Introduction}}
  to {{Thermal Physics}}}}}\ (\bibinfo  {publisher} {{Addison Wesley}},\
  \bibinfo {year} {1999})\BibitemShut {NoStop}%
\bibitem [{\citenamefont {Parrondo}\ \emph {et~al.}(2015)\citenamefont
  {Parrondo}, \citenamefont {Horowitz},\ and\ \citenamefont
  {Sagawa}}]{parrondo_thermodynamics_2015}%
  \BibitemOpen
  \bibfield  {author} {\bibinfo {author} {\bibfnamefont {J.~M.~R.}\
  \bibnamefont {Parrondo}}, \bibinfo {author} {\bibfnamefont {J.~M.}\
  \bibnamefont {Horowitz}},\ and\ \bibinfo {author} {\bibfnamefont
  {T.}~\bibnamefont {Sagawa}},\ }\bibfield  {title} {\bibinfo {title}
  {Thermodynamics of information},\ }\href {https://doi.org/10.1038/nphys3230}
  {\bibfield  {journal} {\bibinfo  {journal} {Nature Physics}\ }\textbf
  {\bibinfo {volume} {11}},\ \bibinfo {pages} {131} (\bibinfo {year}
  {2015})}\BibitemShut {NoStop}%
\bibitem [{\citenamefont {Sagawa}\ and\ \citenamefont
  {Ueda}(2008)}]{sagawa_second_2008}%
  \BibitemOpen
  \bibfield  {author} {\bibinfo {author} {\bibfnamefont {T.}~\bibnamefont
  {Sagawa}}\ and\ \bibinfo {author} {\bibfnamefont {M.}~\bibnamefont {Ueda}},\
  }\bibfield  {title} {\bibinfo {title} {Second {{Law}} of {{Thermodynamics}}
  with {{Discrete Quantum Feedback Control}}},\ }\href
  {https://doi.org/10.1103/PhysRevLett.100.080403} {\bibfield  {journal}
  {\bibinfo  {journal} {Physical Review Letters}\ }\textbf {\bibinfo {volume}
  {100}},\ \bibinfo {pages} {080403} (\bibinfo {year} {2008})}\BibitemShut
  {NoStop}%
\bibitem [{\citenamefont {Kawai}\ \emph {et~al.}(2007)\citenamefont {Kawai},
  \citenamefont {Parrondo},\ and\ \citenamefont {{den
  Broeck}}}]{kawai_dissipation_2007}%
  \BibitemOpen
  \bibfield  {author} {\bibinfo {author} {\bibfnamefont {R.}~\bibnamefont
  {Kawai}}, \bibinfo {author} {\bibfnamefont {J.~M.~R.}\ \bibnamefont
  {Parrondo}},\ and\ \bibinfo {author} {\bibfnamefont {C.~V.}\ \bibnamefont
  {{den Broeck}}},\ }\bibfield  {title} {\bibinfo {title} {Dissipation: {{The
  Phase-Space Perspective}}},\ }\href
  {https://doi.org/10.1103/PhysRevLett.98.080602} {\bibfield  {journal}
  {\bibinfo  {journal} {Physical Review Letters}\ }\textbf {\bibinfo {volume}
  {98}},\ \bibinfo {pages} {080602} (\bibinfo {year} {2007})}\BibitemShut
  {NoStop}%
\bibitem [{\citenamefont {Rubino}\ \emph {et~al.}(2022)\citenamefont {Rubino},
  \citenamefont {Manzano}, \citenamefont {Rozema}, \citenamefont {Walther},
  \citenamefont {Parrondo},\ and\ \citenamefont
  {Brukner}}]{rubino_inferring_2022}%
  \BibitemOpen
  \bibfield  {author} {\bibinfo {author} {\bibfnamefont {G.}~\bibnamefont
  {Rubino}}, \bibinfo {author} {\bibfnamefont {G.}~\bibnamefont {Manzano}},
  \bibinfo {author} {\bibfnamefont {L.~A.}\ \bibnamefont {Rozema}}, \bibinfo
  {author} {\bibfnamefont {P.}~\bibnamefont {Walther}}, \bibinfo {author}
  {\bibfnamefont {J.~M.~R.}\ \bibnamefont {Parrondo}},\ and\ \bibinfo {author}
  {\bibfnamefont {C.}~\bibnamefont {Brukner}},\ }\bibfield  {title} {\bibinfo
  {title} {Inferring work by quantum superposing forward and time-reversal
  evolutions},\ }\href {https://doi.org/10.1103/PhysRevResearch.4.013208}
  {\bibfield  {journal} {\bibinfo  {journal} {Physical Review Research}\
  }\textbf {\bibinfo {volume} {4}},\ \bibinfo {pages} {013208} (\bibinfo {year}
  {2022})}\BibitemShut {NoStop}%
\bibitem [{\citenamefont {Rubino}\ \emph {et~al.}(2021)\citenamefont {Rubino},
  \citenamefont {Manzano},\ and\ \citenamefont
  {Brukner}}]{rubino_quantum_2021}%
  \BibitemOpen
  \bibfield  {author} {\bibinfo {author} {\bibfnamefont {G.}~\bibnamefont
  {Rubino}}, \bibinfo {author} {\bibfnamefont {G.}~\bibnamefont {Manzano}},\
  and\ \bibinfo {author} {\bibfnamefont {C.}~\bibnamefont {Brukner}},\
  }\bibfield  {title} {\bibinfo {title} {Quantum superposition of thermodynamic
  evolutions with opposing time's arrows},\ }\href
  {https://doi.org/10.1038/s42005-021-00759-1} {\bibfield  {journal} {\bibinfo
  {journal} {Communications Physics}\ }\textbf {\bibinfo {volume} {4}},\
  \bibinfo {pages} {1} (\bibinfo {year} {2021})}\BibitemShut {NoStop}%
\bibitem [{\citenamefont {Wald}(2001)}]{wald_thermodynamics_2001}%
  \BibitemOpen
  \bibfield  {author} {\bibinfo {author} {\bibfnamefont {R.~M.}\ \bibnamefont
  {Wald}},\ }\bibfield  {title} {\bibinfo {title} {The {{Thermodynamics}} of
  {{Black Holes}}},\ }\href {https://doi.org/10.12942/lrr-2001-6} {\bibfield
  {journal} {\bibinfo  {journal} {Living Reviews in Relativity}\ }\textbf
  {\bibinfo {volume} {4}},\ \bibinfo {pages} {6} (\bibinfo {year}
  {2001})}\BibitemShut {NoStop}%
\bibitem [{\citenamefont {Bousso}(2002)}]{bousso_holographic_2002}%
  \BibitemOpen
  \bibfield  {author} {\bibinfo {author} {\bibfnamefont {R.}~\bibnamefont
  {Bousso}},\ }\bibfield  {title} {\bibinfo {title} {The holographic
  principle},\ }\href {https://doi.org/10.1103/RevModPhys.74.825} {\bibfield
  {journal} {\bibinfo  {journal} {Reviews of Modern Physics}\ }\textbf
  {\bibinfo {volume} {74}},\ \bibinfo {pages} {825} (\bibinfo {year}
  {2002})}\BibitemShut {NoStop}%
\bibitem [{\citenamefont {Bekenstein}(2004)}]{bekenstein_black_2004}%
  \BibitemOpen
  \bibfield  {author} {\bibinfo {author} {\bibfnamefont {J.~D.}\ \bibnamefont
  {Bekenstein}},\ }\bibfield  {title} {\bibinfo {title} {Black holes and
  information theory},\ }\href {https://doi.org/10.1080/00107510310001632523}
  {\bibfield  {journal} {\bibinfo  {journal} {Contemporary Physics}\ }\textbf
  {\bibinfo {volume} {45}},\ \bibinfo {pages} {31} (\bibinfo {year}
  {2004})}\BibitemShut {NoStop}%
\bibitem [{\citenamefont {Bardeen}\ \emph {et~al.}(1973)\citenamefont
  {Bardeen}, \citenamefont {Carter},\ and\ \citenamefont
  {Hawking}}]{bardeen_four_1973}%
  \BibitemOpen
  \bibfield  {author} {\bibinfo {author} {\bibfnamefont {J.~M.}\ \bibnamefont
  {Bardeen}}, \bibinfo {author} {\bibfnamefont {B.}~\bibnamefont {Carter}},\
  and\ \bibinfo {author} {\bibfnamefont {S.~W.}\ \bibnamefont {Hawking}},\
  }\bibfield  {title} {\bibinfo {title} {The four laws of black hole
  mechanics},\ }\href {https://doi.org/10.1007/BF01645742} {\bibfield
  {journal} {\bibinfo  {journal} {Communications in Mathematical Physics}\
  }\textbf {\bibinfo {volume} {31}},\ \bibinfo {pages} {161} (\bibinfo {year}
  {1973})}\BibitemShut {NoStop}%
\bibitem [{\citenamefont {Hawking}(1975)}]{hawking_particle_1975}%
  \BibitemOpen
  \bibfield  {author} {\bibinfo {author} {\bibfnamefont {S.~W.}\ \bibnamefont
  {Hawking}},\ }\bibfield  {title} {\bibinfo {title} {Particle creation by
  black holes},\ }\href {https://doi.org/10.1007/BF02345020} {\bibfield
  {journal} {\bibinfo  {journal} {Communications in Mathematical Physics}\
  }\textbf {\bibinfo {volume} {43}},\ \bibinfo {pages} {199} (\bibinfo {year}
  {1975})}\BibitemShut {NoStop}%
\bibitem [{\citenamefont {Sekino}\ and\ \citenamefont
  {Susskind}(2008)}]{sekino_fast_2008}%
  \BibitemOpen
  \bibfield  {author} {\bibinfo {author} {\bibfnamefont {Y.}~\bibnamefont
  {Sekino}}\ and\ \bibinfo {author} {\bibfnamefont {L.}~\bibnamefont
  {Susskind}},\ }\bibfield  {title} {\bibinfo {title} {Fast scramblers},\
  }\href {https://doi.org/10.1088/1126-6708/2008/10/065} {\bibfield  {journal}
  {\bibinfo  {journal} {Journal of High Energy Physics}\ }\textbf {\bibinfo
  {volume} {2008}},\ \bibinfo {pages} {065} (\bibinfo {year}
  {2008})}\BibitemShut {NoStop}%
\bibitem [{\citenamefont {Bekenstein}(1981)}]{bekenstein_universal_1981}%
  \BibitemOpen
  \bibfield  {author} {\bibinfo {author} {\bibfnamefont {J.~D.}\ \bibnamefont
  {Bekenstein}},\ }\bibfield  {title} {\bibinfo {title} {Universal upper bound
  on the entropy-to-energy ratio for bounded systems},\ }\href
  {https://doi.org/10.1103/PhysRevD.23.287} {\bibfield  {journal} {\bibinfo
  {journal} {Physical Review D}\ }\textbf {\bibinfo {volume} {23}},\ \bibinfo
  {pages} {287} (\bibinfo {year} {1981})}\BibitemShut {NoStop}%
\bibitem [{\citenamefont {Hubeny}(2015)}]{hubeny_adscft_2015}%
  \BibitemOpen
  \bibfield  {author} {\bibinfo {author} {\bibfnamefont {V.~E.}\ \bibnamefont
  {Hubeny}},\ }\bibfield  {title} {\bibinfo {title} {The {{AdS}}/{{CFT}}
  correspondence},\ }\href {https://doi.org/10.1088/0264-9381/32/12/124010}
  {\bibfield  {journal} {\bibinfo  {journal} {Classical and Quantum Gravity}\
  }\textbf {\bibinfo {volume} {32}},\ \bibinfo {pages} {124010} (\bibinfo
  {year} {2015})}\BibitemShut {NoStop}%
\bibitem [{\citenamefont {Susskind}(1995)}]{susskind_world_1995}%
  \BibitemOpen
  \bibfield  {author} {\bibinfo {author} {\bibfnamefont {L.}~\bibnamefont
  {Susskind}},\ }\bibfield  {title} {\bibinfo {title} {The world as a
  hologram},\ }\href {https://doi.org/10.1063/1.531249} {\bibfield  {journal}
  {\bibinfo  {journal} {Journal of Mathematical Physics}\ }\textbf {\bibinfo
  {volume} {36}},\ \bibinfo {pages} {6377} (\bibinfo {year}
  {1995})}\BibitemShut {NoStop}%
\bibitem [{\citenamefont {Susskind}\ and\ \citenamefont
  {Zhao}(2018)}]{susskind_teleportation_2018}%
  \BibitemOpen
  \bibfield  {author} {\bibinfo {author} {\bibfnamefont {L.}~\bibnamefont
  {Susskind}}\ and\ \bibinfo {author} {\bibfnamefont {Y.}~\bibnamefont
  {Zhao}},\ }\bibfield  {title} {\bibinfo {title} {Teleportation {{Through}}
  the {{Wormhole}}},\ }\href {https://doi.org/10.1103/PhysRevD.98.046016}
  {\bibfield  {journal} {\bibinfo  {journal} {Physical Review D}\ }\textbf
  {\bibinfo {volume} {98}},\ \bibinfo {pages} {046016} (\bibinfo {year}
  {2018})},\ \Eprint {https://arxiv.org/abs/1707.04354} {arXiv:1707.04354
  [gr-qc, physics:hep-th, physics:quant-ph]} \BibitemShut {NoStop}%
\bibitem [{\citenamefont {Ryu}\ and\ \citenamefont
  {Takayanagi}(2006)}]{ryu_aspects_2006}%
  \BibitemOpen
  \bibfield  {author} {\bibinfo {author} {\bibfnamefont {S.}~\bibnamefont
  {Ryu}}\ and\ \bibinfo {author} {\bibfnamefont {T.}~\bibnamefont
  {Takayanagi}},\ }\bibfield  {title} {\bibinfo {title} {Aspects of holographic
  entanglement entropy},\ }\href
  {https://doi.org/10.1088/1126-6708/2006/08/045} {\bibfield  {journal}
  {\bibinfo  {journal} {Journal of High Energy Physics}\ }\textbf {\bibinfo
  {volume} {2006}},\ \bibinfo {pages} {045} (\bibinfo {year}
  {2006})}\BibitemShut {NoStop}%
\bibitem [{\citenamefont {Moustos}(2017)}]{moustos_gravity_2017}%
  \BibitemOpen
  \bibfield  {author} {\bibinfo {author} {\bibfnamefont {D.}~\bibnamefont
  {Moustos}},\ }\href {https://doi.org/10.48550/arXiv.1701.08967} {\bibinfo
  {title} {Gravity as a thermodynamic phenomenon}} (\bibinfo {year} {2017}),\
  \Eprint {https://arxiv.org/abs/1701.08967} {arXiv:1701.08967 [gr-qc]}
  \BibitemShut {NoStop}%
\bibitem [{\citenamefont {Ashby}(1991)}]{ashby_requisite_1991}%
  \BibitemOpen
  \bibfield  {author} {\bibinfo {author} {\bibfnamefont {W.~R.}\ \bibnamefont
  {Ashby}},\ }\bibfield  {title} {\bibinfo {title} {Requisite {{Variety}} and
  {{Its Implications}} for the {{Control}} of {{Complex Systems}}},\ }in\ \href
  {https://doi.org/10.1007/978-1-4899-0718-9_28} {\emph {\bibinfo {booktitle}
  {Facets of {{Systems Science}}}}},\ \bibinfo {series and number}
  {International {{Federation}} for {{Systems Research International Series}}
  on {{Systems Science}} and {{Engineering}}},\ \bibinfo {editor} {edited by\
  \bibinfo {editor} {\bibfnamefont {G.~J.}\ \bibnamefont {Klir}}}\ (\bibinfo
  {publisher} {{Springer US}},\ \bibinfo {address} {{Boston, MA}},\ \bibinfo
  {year} {1991})\ pp.\ \bibinfo {pages} {405--417}\BibitemShut {NoStop}%
\bibitem [{\citenamefont {Brown}\ \emph {et~al.}(2016)\citenamefont {Brown},
  \citenamefont {Roberts}, \citenamefont {Susskind}, \citenamefont {Swingle},\
  and\ \citenamefont {Zhao}}]{brown_complexity_2016}%
  \BibitemOpen
  \bibfield  {author} {\bibinfo {author} {\bibfnamefont {A.~R.}\ \bibnamefont
  {Brown}}, \bibinfo {author} {\bibfnamefont {D.~A.}\ \bibnamefont {Roberts}},
  \bibinfo {author} {\bibfnamefont {L.}~\bibnamefont {Susskind}}, \bibinfo
  {author} {\bibfnamefont {B.}~\bibnamefont {Swingle}},\ and\ \bibinfo {author}
  {\bibfnamefont {Y.}~\bibnamefont {Zhao}},\ }\bibfield  {title} {\bibinfo
  {title} {Complexity, action, and black holes},\ }\href
  {https://doi.org/10.1103/PhysRevD.93.086006} {\bibfield  {journal} {\bibinfo
  {journal} {Physical Review D}\ }\textbf {\bibinfo {volume} {93}},\ \bibinfo
  {pages} {086006} (\bibinfo {year} {2016})}\BibitemShut {NoStop}%
\bibitem [{\citenamefont {Bouland}\ \emph {et~al.}(2019)\citenamefont
  {Bouland}, \citenamefont {Fefferman},\ and\ \citenamefont
  {Vazirani}}]{bouland_computational_2019}%
  \BibitemOpen
  \bibfield  {author} {\bibinfo {author} {\bibfnamefont {A.}~\bibnamefont
  {Bouland}}, \bibinfo {author} {\bibfnamefont {B.}~\bibnamefont {Fefferman}},\
  and\ \bibinfo {author} {\bibfnamefont {U.}~\bibnamefont {Vazirani}},\ }\href
  {https://doi.org/10.48550/arXiv.1910.14646} {\bibinfo {title} {Computational
  pseudorandomness, the wormhole growth paradox, and constraints on the
  {{AdS}}/{{CFT}} duality}} (\bibinfo {year} {2019}),\ \Eprint
  {https://arxiv.org/abs/1910.14646} {arXiv:1910.14646 [gr-qc, physics:hep-th,
  physics:quant-ph]} \BibitemShut {NoStop}%
\bibitem [{\citenamefont {Crooks}(1999)}]{crooks_entropy_1999}%
  \BibitemOpen
  \bibfield  {author} {\bibinfo {author} {\bibfnamefont {G.~E.}\ \bibnamefont
  {Crooks}},\ }\bibfield  {title} {\bibinfo {title} {Entropy production
  fluctuation theorem and the nonequilibrium work relation for free energy
  differences},\ }\href {https://doi.org/10.1103/PhysRevE.60.2721} {\bibfield
  {journal} {\bibinfo  {journal} {Physical Review E}\ }\textbf {\bibinfo
  {volume} {60}},\ \bibinfo {pages} {2721} (\bibinfo {year}
  {1999})}\BibitemShut {NoStop}%
\bibitem [{\citenamefont {Evans}\ and\ \citenamefont
  {Searles}(2002)}]{evans_fluctuation_2002}%
  \BibitemOpen
  \bibfield  {author} {\bibinfo {author} {\bibfnamefont {D.~J.}\ \bibnamefont
  {Evans}}\ and\ \bibinfo {author} {\bibfnamefont {D.~J.}\ \bibnamefont
  {Searles}},\ }\bibfield  {title} {\bibinfo {title} {The {{Fluctuation
  Theorem}}},\ }\href {https://doi.org/10.1080/00018730210155133} {\bibfield
  {journal} {\bibinfo  {journal} {Advances in Physics}\ }\textbf {\bibinfo
  {volume} {51}},\ \bibinfo {pages} {1529} (\bibinfo {year}
  {2002})}\BibitemShut {NoStop}%
\bibitem [{\citenamefont {Lodato}\ and\ \citenamefont
  {Natarajan}(2006)}]{lodato_supermassive_2006}%
  \BibitemOpen
  \bibfield  {author} {\bibinfo {author} {\bibfnamefont {G.}~\bibnamefont
  {Lodato}}\ and\ \bibinfo {author} {\bibfnamefont {P.}~\bibnamefont
  {Natarajan}},\ }\bibfield  {title} {\bibinfo {title} {Supermassive black hole
  formation during the assembly of pre-galactic discs},\ }\href
  {https://doi.org/10.1111/j.1365-2966.2006.10801.x} {\bibfield  {journal}
  {\bibinfo  {journal} {Monthly Notices of the Royal Astronomical Society}\
  }\textbf {\bibinfo {volume} {371}},\ \bibinfo {pages} {1813} (\bibinfo {year}
  {2006})}\BibitemShut {NoStop}%
\bibitem [{\citenamefont {Hawking}(1974)}]{hawking_black_1974}%
  \BibitemOpen
  \bibfield  {author} {\bibinfo {author} {\bibfnamefont {S.~W.}\ \bibnamefont
  {Hawking}},\ }\bibfield  {title} {\bibinfo {title} {Black hole explosions?},\
  }\href {https://doi.org/10.1038/248030a0} {\bibfield  {journal} {\bibinfo
  {journal} {Nature}\ }\textbf {\bibinfo {volume} {248}},\ \bibinfo {pages}
  {30} (\bibinfo {year} {1974})}\BibitemShut {NoStop}%
\bibitem [{\citenamefont {Andrieux}\ and\ \citenamefont
  {Gaspard}(2006)}]{andrieux_fluctuation_2006}%
  \BibitemOpen
  \bibfield  {author} {\bibinfo {author} {\bibfnamefont {D.}~\bibnamefont
  {Andrieux}}\ and\ \bibinfo {author} {\bibfnamefont {P.}~\bibnamefont
  {Gaspard}},\ }\bibfield  {title} {\bibinfo {title} {Fluctuation theorem for
  transport in mesoscopic systems},\ }\href
  {https://doi.org/10.1088/1742-5468/2006/01/P01011} {\bibfield  {journal}
  {\bibinfo  {journal} {Journal of Statistical Mechanics: Theory and
  Experiment}\ }\textbf {\bibinfo {volume} {2006}},\ \bibinfo {pages} {P01011}
  (\bibinfo {year} {2006})}\BibitemShut {NoStop}%
\bibitem [{\citenamefont {Esposito}\ and\ \citenamefont
  {Mukamel}(2006)}]{esposito_fluctuation_2006}%
  \BibitemOpen
  \bibfield  {author} {\bibinfo {author} {\bibfnamefont {M.}~\bibnamefont
  {Esposito}}\ and\ \bibinfo {author} {\bibfnamefont {S.}~\bibnamefont
  {Mukamel}},\ }\bibfield  {title} {\bibinfo {title} {Fluctuation theorems for
  quantum master equations},\ }\href
  {https://doi.org/10.1103/PhysRevE.73.046129} {\bibfield  {journal} {\bibinfo
  {journal} {Physical Review E}\ }\textbf {\bibinfo {volume} {73}},\ \bibinfo
  {pages} {046129} (\bibinfo {year} {2006})}\BibitemShut {NoStop}%
\bibitem [{\citenamefont {Parikh}\ and\ \citenamefont
  {Wilczek}(2000)}]{parikh_hawking_2000}%
  \BibitemOpen
  \bibfield  {author} {\bibinfo {author} {\bibfnamefont {M.~K.}\ \bibnamefont
  {Parikh}}\ and\ \bibinfo {author} {\bibfnamefont {F.}~\bibnamefont
  {Wilczek}},\ }\bibfield  {title} {\bibinfo {title} {Hawking {{Radiation As
  Tunneling}}},\ }\href {https://doi.org/10.1103/PhysRevLett.85.5042}
  {\bibfield  {journal} {\bibinfo  {journal} {Physical Review Letters}\
  }\textbf {\bibinfo {volume} {85}},\ \bibinfo {pages} {5042} (\bibinfo {year}
  {2000})}\BibitemShut {NoStop}%
\bibitem [{\citenamefont {Rosa}(2010)}]{rosa_extremal_2010}%
  \BibitemOpen
  \bibfield  {author} {\bibinfo {author} {\bibfnamefont {J.}~\bibnamefont
  {Rosa}},\ }\bibfield  {title} {\bibinfo {title} {The extremal black hole
  bomb},\ }\href {https://doi.org/10.1007/JHEP06(2010)015} {\bibfield
  {journal} {\bibinfo  {journal} {Journal of High Energy Physics}\ }\textbf
  {\bibinfo {volume} {2010}},\ \bibinfo {pages} {15} (\bibinfo {year}
  {2010})}\BibitemShut {NoStop}%
\end{thebibliography}%

\end{document}